\documentclass[conference]{IEEEtran}
\IEEEoverridecommandlockouts
\usepackage{cite}
\usepackage{amsmath,amssymb,amsfonts}
\usepackage{graphicx}
\usepackage{textcomp}
\usepackage[dvipsnames]{xcolor}
\usepackage{url}
\usepackage{multirow}
\usepackage{hyperref}
\usepackage{longtable}
\usepackage{tabu}
\usepackage{colortbl}
\usepackage[ruled,linesnumbered]{algorithm2e}
\usepackage{wrapfig}
\usepackage{tcolorbox}

\def\BibTeX{{\rm B\kern-.05em{\sc i\kern-.025em b}\kern-.08em
    T\kern-.1667em\lower.7ex\hbox{E}\kern-.125emX}}
\begin{document}

\title{Label Flipping Data Poisoning Attack Against Wearable Human Activity Recognition System}

\author{\IEEEauthorblockN{Abdur R. Shahid$^1$, Ahmed Imteaj$^2$, Peter Y. Wu$^1$, Diane A. Igoche$^1$, Tauhidul Alam$^3$}

\IEEEauthorblockA{\textit{$^{1}$Department of Computer and Information Systems, Robert Morris University, Moon, PA, USA} \\
\textit{$^{2}$School of Computing, Southern Illinois University, Carbondale, IL, USA} \\
\textit{$^{3}$Department of Computer Science, Louisiana State University Shreveport, Shreveport, LA, USA}\\
$^{1}$\{shahid, wu, igoche\}@rmu.edu, 
$^{2}$aimteaj@gmail.com, talam@lsus.edu}
}

\maketitle

\begin{abstract}
Human Activity Recognition (HAR) is a problem of interpreting sensor data to human movement using an efficient machine learning (ML) approach. The HAR systems rely on data from untrusted users, making them susceptible to data poisoning attacks. In a poisoning attack, attackers manipulate the sensor readings to contaminate the training set, misleading the HAR to produce erroneous outcomes. This paper presents the design of a label flipping data poisoning attack for a HAR system, where the label of a sensor reading is maliciously changed in the data collection phase. Due to high noise and uncertainty in the sensing environment, such an attack poses a severe threat to the recognition system. Besides, vulnerability to label flipping attacks is dangerous when activity recognition models are deployed in safety-critical applications. This paper shades light on how to carry out the attack in practice through smartphone-based sensor data collection applications. This is an earlier research work, to our knowledge, that explores attacking the HAR models via label flipping poisoning. We implement the proposed attack and test it on activity recognition models based on the following machine learning algorithms: multi-layer perceptron, decision tree, random forest, and XGBoost. Finally, we evaluate the effectiveness of a K-nearest neighbors (KNN)-based defense mechanism against the proposed attack.
\end{abstract}


\section{Introduction}
Human Activity Recognition, or HAR, is a field of study to understand, recognize, and predict human activities through machine learning (ML). In HAR, raw time-series data from wearable sensors are acquired and translated to human activities (e.g. walking, sitting, jumping, fighting, falling, etc.).
Thanks to the ubiquity of wearable sensors and the Internet of Things (IoT) and their ever-growing computing, networking, and sensing powers, designing and building effective and efficient wearable-sensor-based human activity recognition (HAR) systems have gained increased attention from researchers and tech industries in the last decade. Apart from the various wearable sensor-based HAR systems proposed in the literature, various commercial wearable products are now available on the market including fitness trackers, smartwatches, and smartphones which are packed with various wearable sensors. Commonly utilized wearable sensors in HAR systems\cite{dang2020sensor} include accelerometer\cite{johnson2020multidimensional}, magnetometer, gyroscopes, inertial measurement unit (IMU), electromyogram (EMG)\cite{al2020effective}, force-sensitive resistors (FSR)\cite{dontha2022wearable} and wearable wrist camera\cite{chen2018wristcam}. 
The wearable sensors-driven HAR has critical applications in healthcare \cite{wang2019survey, matsushita2021recent, palaniappan2012abnormal, kristoffersson2022systematic}, human-robot interaction \cite{lee2019model, tang2022multi}, interactive gaming \cite{dontha2022wearable}, sports \cite{camomilla2018trends, johnson2020multidimensional}, military \cite{kang2020no, kang2021military} and so on. 

Wearable-sensor-based HAR is a classification problem that involves data acquisition, preprocessing to filter out signal variability or noise, feature extraction, training, and validating the recognizer. Similar to other ML problems, the success of HAR largely depends on the quality and quantity of the dataset to train and develop the recognition model. While over the last decades ML-based solutions have achieved quite a success, the adversarial side of data collection for HAR is yet to be explored. To understand the magnitude of the problem, we need to look at the two modes of data collection for HAR: closed environment and mobile crowdsensing (MCS). In a closed environment, the data collection is done in a controlled way and hence the risk of adversarial attacks is close to zero. However, on the downside, this limits the variety and size of the dataset, which can cause sampling bias in the dataset. For instance, the UCI HAR dataset consists of data of 30 volunteers \cite{anguita2013public}. Another popular HAR dataset, WISDM, was acquired from 51 test subjects \cite{weiss2019smartphone}. The MHealth dataset contains data from ten volunteers \cite{banos2014mhealthdroid}. Such limitation is a real hindrance to the development process of a HAR system. While MCS has the great potential to solve this problem, its openness allows malicious entities to attack the system easily. In MCS, adversaries have the privilege of sending corrupted samples directly to the dataset aggregator to poison the recognition system; and hence mislead to produce erroneous outcomes. Such an attack is referred to as a data poisoning attack. To develop a robust HAR system, it is critical to study how well a HAR system performs under poisoning attacks. In a training-only data poisoning attack, the attacker is capable of manipulating the training instances of the target model without the need of accessing the test instances. Several training-only data poisoning attacks have been proposed in literature including feature collision, influence functions, label flipping, vanishing gradients, generative models, and model poisoning\cite{goldblum2022dataset}. As the classifier learns from poisoned data, it will lead to unintentional or even life-threaten situations. Let us consider a HAR system designed to call for emergency service if detects a sudden fall. For fun, profit, and revenge, attackers might attempt to alter the training dataset so that the trained model on the poisoned dataset recognizes a heart attack-related fall as a regular sitting. This is an example of \textit{targeted attack} in which the attacker's goal is to change the behavior of the model on particular instances. In \textit{untargeted attacks}, attackers seek to alter a model's behavior regardless of particular instances or scenarios. A subclass of data poisoning attack is \textit{label flipping attack} where the label of a sensor reading is maliciously changed in the data collection phase\cite{biggio2011support,xiao2012adversarial, paudice2018label,baracaldo2018detecting}. Let us consider a HAR system to recognize the following activities using a smartphone: \{walking, going upstairs, going downstairs, sitting, standing, laying, crawling\}. In an untargeted label flipping attack, an attacker might aim to jeopardize the model development process by injecting label flipped data as much as possible. To illustrate, instead of sending data for walking, the attacker might change the label to standing. 

In this work, we focus on untargeted label flipping attack that can happen through mobile crowdsensing. While there has been a long line of prior work on data poisoning, very few of them covered HAR systems. In fact, data poisoning attack on the IoT system is less studied compared to vision-related system. In this paper, we make the following contributions.
\begin{itemize}
\item
\textbf{Label Flipping Attack for Human Activity Recognition Systems.}
We present a label flipping attack for multi-class sensor-based HAR systems. To the best of our knowledge, this is the first work on label flipping attacks for sensor-based HAR systems. We first propose an optimal but computationally intractable version of the problem. Next, to reduce the computational cost, we present a randomized approach for the attack.

\item 
\textbf{Label Flipping Attack for Multi-Class Classification Task.}
While most contemporary works on label flipping attacks focus on binary classification problems, we extend the existing works to design the attack for multi-class classifiers.

%\item
%\textbf{Practical Approach for Label Flipping Attack on User Side.}
%We explore sensor various smartphone apps from the Google Play store which can be practically used to collect data from various sensors. Then, we demonstrate how easily the label flipping attack can be performed using a smartphone application.

\item
\textbf{k-NN-based Defense Mechanism.} We extend K-Nearest Neighbor (KNN)-based defense mechanism, proposed by Paudice et al.\cite{paudice2018label}, to evaluate the proposed attack. This extended mechanism is capable of detecting the malicious training data (whose label was changed) and predicting their correct label. The mechanism is suitable for participatory-MCS, where the collected data is partially trusted. 

\item
\textbf{Empirical Study of the Label Flipping Attack and Defense.} 
We empirically evaluate the effectiveness of the proposed attack and defense mechanisms. In the evaluation, we utilized the popular UCI HAR dataset\cite{anguita2013public} and developed HAR models using the following algorithms: Multi-layer Perceptron, Decision Tree, Random Forest, and XGBoost in a blackbox setting.   
\end{itemize}

\begin{figure}
\centering
\includegraphics[width=.85\linewidth]{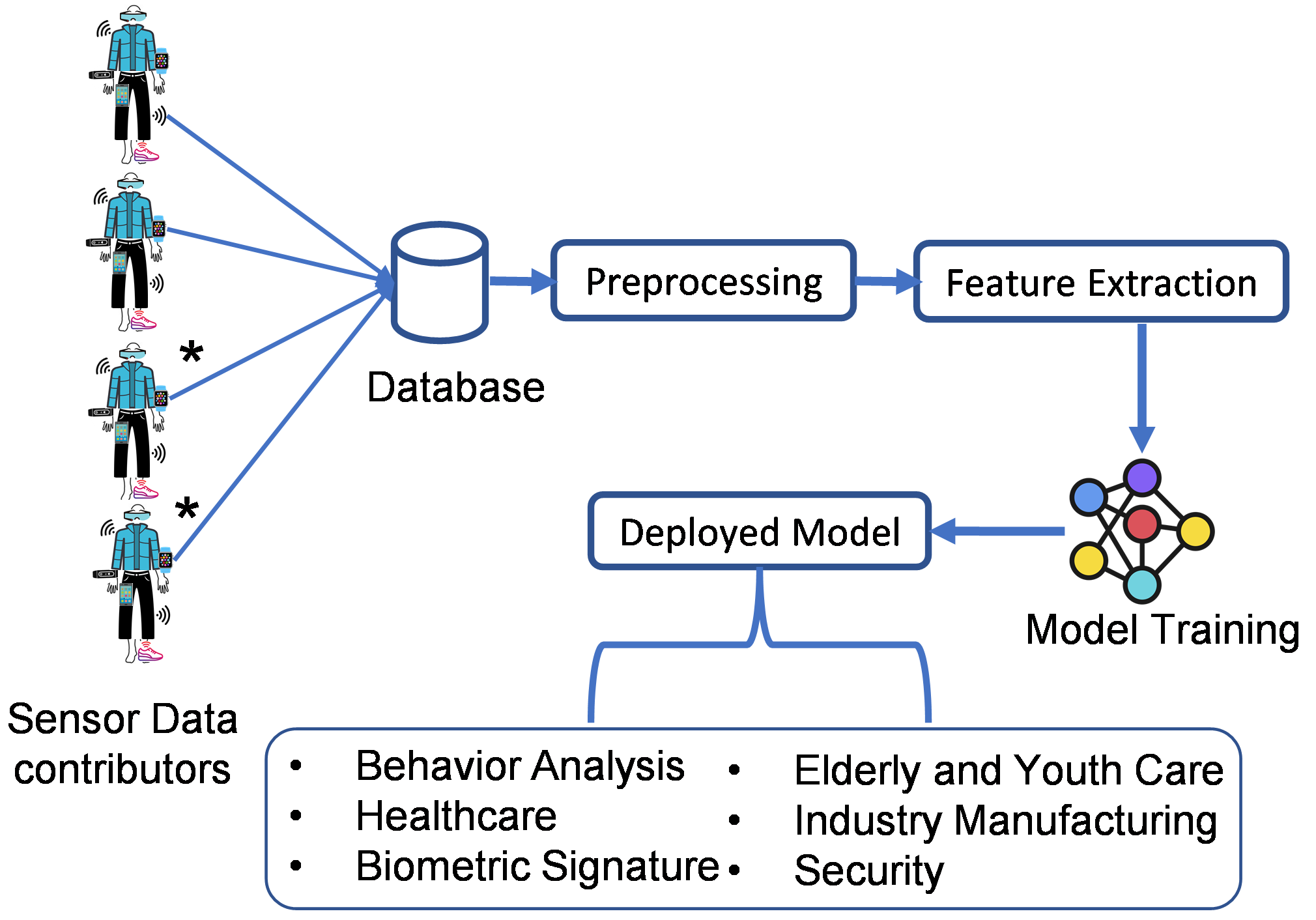}
\caption{HAR system with sensor data contributors. $\star$ depicts attackers who submit label flipped poisonous data to the system.\label{fig:motivation}}
\end{figure}

\begin{table} 
\caption{Frequently used Symbols \label{tab:symbols}}

    \begin{tabular}{l|l}
    \textbf{Symbol}    &   \textbf{Definition} \\  \hline
    $\mathcal{D}$ & Dataset \\  
    $\beta$ & Threshold for number of subsets \\  
    $\theta$ & Threshold for loss function output \\
    $x$ & Trusted dataset size \\
    $p$ & Number of label flipped samples \\
    $K$ & Defense model's hyperparameter 
    
    \end{tabular}

\end{table}

\section{Related Work}

Poisoning attack against machine learning algorithms has become an emerging research in the field of adversarial machine
learning \cite{barreno2006can, goldblum2022dataset, pmlr-v9-kloft10a}. Over the years, different types of poisoning attacks have been proposed in the literature, including label flipping attack \cite{biggio2011support,xiao2012adversarial, paudice2018label,baracaldo2018detecting}. These attacks work by switching training labels while leaving the data instances untouched.  They have the advantage of not introducing strange-looking artifacts, which may be obvious to the intended victim. One prominent work in this area is the attack proposed by Biggio et al.\cite{biggio2011support} to evaluate the effectiveness of Support Vector Machines (SVM) in adversarial classification tasks. The authors adopt two different strategies for contaminating the training set through label flipping: random and adversarial label flips. In both cases, they assumed that an adversary can only flip the labels of a given percentage of training samples. Zhang et al.\cite{zhang2017game} established a game-theoretic framework for attacking SVM models. Zhao et al.\cite{zhao2017efficient} developed Projected Gradient Ascent
(PGA) algorithm to compute a label contamination attack (LCA) which improved the SVM-based works. paudice et al.\cite{paudice2018label} proposed an efficient algorithm to perform
optimal label flipping poisoning attacks. 
From the perspective of defending against label flipping poisoning attacks on modern ML, Paudice et al.\cite{paudice2018label} proposed k-Nearest-Neighbors (kNN) to re-label reach data point to mitigate label flipping attacks. Specifically, they re-label each data point with the most common label among its $K$ nearest neighbors.

A major limitation of most of the works on designing label flipping work concentrated on binary classification problems. Also, an investigation of the effectiveness of the proposed attacks on wearable-based HAR is yet to be explored. 

\section{The Setting}
\subsection{Classification Problem}
Consider a multi-class human-activity classification problem: a recognizer ($f:\mathcal{X}\rightarrow \mathcal{R}^k$) receives a pair of random variables $(X,Y)\in \mathcal{X}\times \{1,\ldots k\}$ where $Y$ is unobserved and wishes to assign the variable $X$ to one of the $k$ classes $ K = \{1, 2,\ldots k\}$ such that the probability of a misclassification is minimized. Each component $f_y(x), y = 1,\ldots x$ is the deemed likelihood the recognizer assigns to class $y$ for $x$. Consequently, the goal of the recognizer is then to minimize the expected loss function $L$,
\begin{equation}\label{eq:classification}
    R_L(f):=\mathbb{E}[L(f(X), Y)] 
\end{equation}
Where $L(f(X), Y)$ measures the loss of margins $f(x)\in \mathcal{R}^k$ when the true label of $x$ is $y$ and the expectation (eq. \ref{eq:classification}) is taken jointly over $(X, Y)$.

\subsection{Threat Model}
Similar to prior works on data poisoning, we consider a white-box attacker. We consider the worst-case scenario of an attack, where the attacker has the capability of directly injecting data into the training dataset. The attacker has access to the raw data collection phase and is capable of modifying the data to perform an attack. The attacker can disguise itself as an ordinary sensor data contributor. It has knowledge of the data distribution, a portion of the training dataset, the training algorithm, the model type, and hyperparameters. While the consideration of such an attacker seems a bit unrealistic but it allows us to investigate the worst-case performance and demonstrate how robust the learning algorithm is; under certain attacks. Using such knowledge, the goal of the attacker is to manipulate the data collection phase to poison the training data set. 
For this purpose, the attacker replaces the label of its dataset before the model is in the training step. This means, its goal is the opposite of the recognizer: maximize the expected loss $L$. We assume that any defense mechanism deployed by the system to prevent data poisoning is unknown to the attacker.

%attacker's goal is to reduce the overall accuracy of the recognizer.  

\section{The Label Flipping Attack on HAR}

We assumed that the main purpose of the attacker
is to manipulate the labels of some samples in the training set to mislead the model to make wrong predictions as many as possible. In other words, in this problem, the loss function on the true data is minimized and the cost function of the poisoned data is maximized. However, an arbitrary amount of changes might be noticed by the system and subsequently stops the attacker from causing further damage to the model. Therefore, the attack becomes a multi-objective optimization problem where, on one hand, the attacker wants to cause the highest amount of damage to the model, and on the other hand, it wants to avoid any defense mechanism as much as possible. If $\mathcal{D^\text{true}}$, $\mathcal{D^\text{poisoned}}$, and $\mathcal{D}$ are the true/not poisoned, poisoned, and entire sets where $\mathcal{D} = \mathcal{D^\text{true}} \cup \mathcal{D^\text{poisoned}}$; then the attack can be formulated as an optimization problem and can be expressed as follows:
\begin{equation}\label{eq:optim1}
    \arg \min \Big( \sum_{x_i\in \mathcal{D^\text{true}}} L(f(x_i), y_i)) - \sum_{x_i\in \mathcal{D^\text{poisoned}}} L(f(x_i), y_i))\Big)
\end{equation}
To carry out the flipping attack, an adversary targets to choose $p$ samples whose label will be flipped such that the loss function of the model, trained by the attacker, is maximized while the visibility of the attack to the system is minimized. The amount of visibility can be expressed as a budget $\mathcal{C}$. If the cost of flipping a sample's label is $c_i$, then the optimization problem (eq. \ref{eq:optim1}) becomes,

\begin{equation}\label{eq:optim2}
\begin{split}
    \arg \min_{\mathcal{C}} & \Big( \sum_{x_i\in \mathcal{D^\text{true}}} L(f(x_i), y_i)) - \sum_{x_i\in \mathcal{D^\text{poisoned}}} L(f(x_i), y_i))\Big)\\
    & s.t.\quad \sum_{i=1}^{i=p}c_i \leq \mathcal{C}
\end{split}
\end{equation}
We can express the optimal label flipping strategy as follows. In every subset $s \in\left( \begin{array}{c} n \\ p \end{array} \right)$ of $\mathcal{D}$, change the original label of each sample $(X, Y) \in s$ to all the other labels in $(K\setminus Y)$ and generate a   set with contaminated samples. Here, $n$ refers to the number of samples in $\mathcal{D}$. For each of the contaminated sets $\mathcal{D}^\prime$ train a model. Finally, select the training set for which its corresponding model's loss function's output was the maximum among all the contaminated models while it meets specific budget constraints. Solving such a problem requires a heuristic search considering all possible subsets to identify the labels that are flipped. Therefore, next, we present a realistic randomized approach for the label flipping attack. Before that, we need to keep in mind that the design of budget $\mathcal{C}$ depends on various factors, including time and resource constraints. In this work, we propose a version of the attack that can be performed straight from a wearable (e.g.smartphone) during the data collection phase. 

We first introduce two parameters $\beta$ and $\delta$ where $\beta$ is to control the number of subsets to try out of the total $\left( \begin{array}{c} n \\ p \end{array} \right)$ from $S_A$ set. Theoretically, $0<\beta\leq \left( \begin{array}{c} n \\ p \end{array} \right)$. The parameter $\delta$ is to define the attacker's satisfactory level of loss function's output. If the loss function's output meets or exceeds $\delta$ then we say the attacker's goal is achieved. Utilizing these two parameters, the proposed randomized label flipping attack stops its search to find a suboptimal subset if the following condition is true: 
\begin{equation}
\begin{split}
    \beta \geq \text{(Total number of subsets considered)} \text{ and } \\ \delta\geq \text{(The output of the loss function)}
\end{split}
\end{equation}
The method is outlined in algorithm \ref{alg:algo}. This algorithm takes as input:
\begin{enumerate}
    \item A supervised machine learning algorithm
    \item A dataset $\mathcal{D}$
    \item The values of the parameters $\beta$ and $\delta$
\end{enumerate}
Using these inputs, the method first randomly picks $p$ samples from $S_A$ (line 3 of algorithm \ref{alg:algo}). For each selected sample, it then replaces the label with a randomly selected label from the set $K\setminus Y$. As an example, consider a HAR system to recognize the following set of activities:\{walking, sitting, standing, and laying\}. If a sample has the label "laying", the attack method will pick one label from the set \{walking, sitting, standin\} at random, and replace "laying" with the selected label. After flipping the labels of the selected $p$ samples, we get a contaminated set $S^\prime_A$ which is then used to train and validate a contaminated model (lines 4 and 5 of algorithm \ref{alg:algo}). Finally, the method selects that contaminated dataset for which the loss function's output is the maximum (line 10 of algorithm \ref{alg:algo}).

\begin{algorithm}
\caption{Randomized Label Flipping Attack}\label{alg:algo}
%\KwData{$S_A$, $p$, the size $n$ of $S_A$, a supervised machine learning algorithm, and loss function}
%\KwResult{Label flipped data}
$N_s\leftarrow 0, L_o\leftarrow 0$

\While{$N_s\leq \beta$ and $L_o\leq \delta$}{
    Randomly select $p$ samples from $\mathcal{D}$.
    
    Randomly change the label of each of the selected samples and create a contaminated set $\mathcal{D}^\prime$.
    
    $L_o\leftarrow $Train a contaminated model and calculate the loss function of the trained model.
    
    save the model and the contaminated set.
    
    $N_s\leftarrow N_s+1$
}
Select and return the contaminated dataset for which the model's loss function was maximized.

\end{algorithm}

\section{Defense Mechanism}
\begin{figure}
\centering
\includegraphics[width=.85\linewidth]{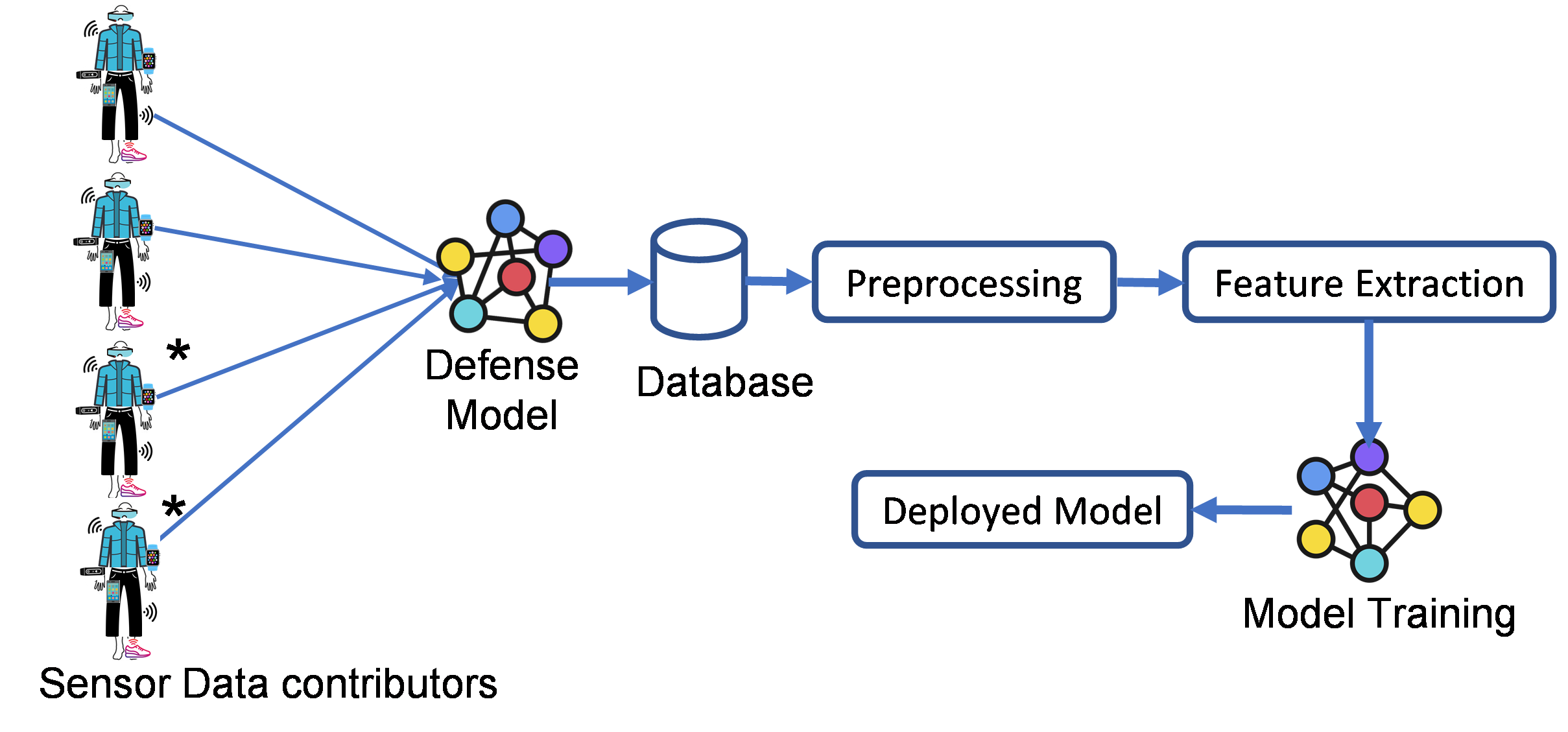}
\caption{HAR system model with defense against the label flipping poisoning attack.\label{fig:defense}}
\end{figure}

For a wearable-based HAR system, it is not difficult to collect some data under a controlled environment. For instance, the data collection can be video recorded and any inconsistency in the label of the collected data from wearable sensors can be corrected. This type of data can be called ``trusted". This notion of the trusted dataset is the central idea of our defense mechanism. Put it another way, our data collection process is a hybrid model covering both a closed environment and crowdsensing. Naturally, the size of an available trusted dataset is smaller than the untrusted dataset. The defense mechanism works as follows. At the beginning, it takes a trusted dataset and a machine learning algorithm as input. Here, we use the KNN approach proposed by Paudice et al.\cite{paudice2018label}. Using the inputs, the mechanism trains a model to predict the label of a given sample. This model is then deployed between the data source and the database to filter out and correct the labels of poisoned samples before inserting them into the database. The design of the system with the defense mechanism is depicted using figure \ref{fig:defense}. Algorithm \ref{alg:defense} describes the steps of the defense mechanism. With $K$ as its hyperparameter, it takes a potential poisonous sample $(x, y)$ as input. Then, it calculates the distance of each sample in the trusted set to the input sample, then pick the $K$ closest samples (line 1 in algorithm \ref{alg:defense}). Finally, it replaces $y$ in $(x,y)$ with the mode of the labels of selected $K$ samples.

\begin{algorithm}
\caption{Label Sanitization}\label{alg:defense}
\KwData{$K$, potentially poisoned sample $(x,y)$, trusted dataset $\mathcal{D^\text{trusted}}$}
\KwResult{Sanitized sample}

Pick the $K$ closest samples from $\mathcal{D^\text{trusted}}$ based on their distances from the poisoned sample.

$y_\text{pred}\leftarrow$ Mode of the labels of selected $K$ samples.

Return $(x,y_\text{pred})$ after replacing $y$ with $y_\text{pred}$ in $(x,y)$.

\end{algorithm}

%%%%%%%%%%%%%%%%%%%%%%%%%%%%%%%%%%%%%%%%%%

%%%%%%%%%%%%%%%%%%%%%%%%%%%%%%%%%%%%%%%%%%

\section{Results}
\subsection{Experimental Setting}
\begin{wrapfigure}{r}{0.22\textwidth}
  \begin{center}
  \vspace{-10pt}
    \includegraphics[width=0.22\textwidth]{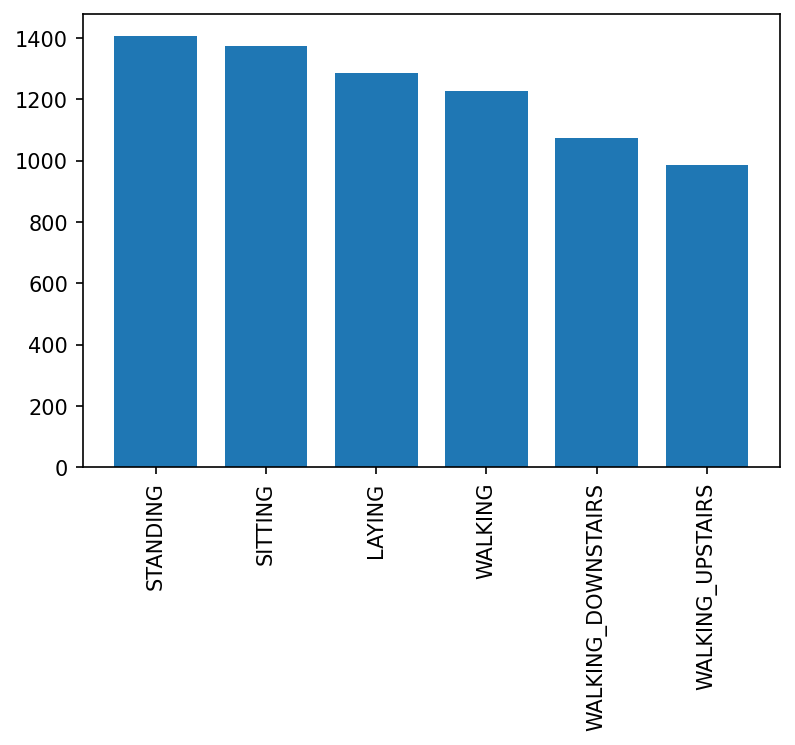}
  \end{center}
  \label{fig:distribution}
  \vspace{-20pt}
  \caption{Distribution of the activities in the dataset.}
  %\hspace{-20pt}
\end{wrapfigure}

In the experiment, we utilize the UCI HAR dataset\cite{anguita2013public}. It consists data of 30 volunteers within an age bracket of 19-48 years. Each person performed six activities (WALKING, WALKING\_UPSTAIRS, WALKING\_DOWNSTAIRS, SITTING, STANDING, LAYING) wearing a smartphone (figure \ref{fig:distribution} shows the distribution of the activities in the dataset).
The experiments were video-recorded to label the data manually. The obtained dataset had been randomly partitioned into two sets, where 75\% of the volunteers were selected for generating the training data and 25\% for the test data. We develop HAR models in both non-adversarial and adversarial environments using the following algorithms:
\begin{itemize}
    \item Multi-layer Perceptron (MLP)
    \item Decision Tree (DT)
    \item Random Forest (RF)
    \item XGBoost 
\end{itemize}
We train all the models in a black-box settings with default parameters provided by scikit-learn package. In the non-adversarial environment, we assume that all the samples are with true labels. In the adversarial environment, we separate $x\%$ samples from the training dataset as \textit{trusted dataset}. This $x\%$ samples are then used to develop a KNN model for defense mechanism. Out of the $(1-x)\%$ training samples, we flip the labels of $p\%$ samples using the proposed attack (algorithm \ref{alg:algo}). For speedy proof of concept generation, we use $\beta = 1$ and $\delta = 0$. Also, such a setting of these two parameters makes resource-constrained attack a reality. The values of $x$, $p$, and $K$ used in the experiment are as follows.
\begin{itemize}
    \item $x$ = $\{5, 10, 15, 20, 25\}$ 
    \item $p$ = $\{1, 5, 10, 15, 20, 25, 30\}$ 
    \item $K$ = $\{3, 5, 7, 9, 11, 13\}$
\end{itemize}
Using the combinations of the values of these parameters, we developed multiple contaminated and defense models.

\begin{table} 
\centering
\caption{Accuracy of the baseline models on the original dataset (in \%)\label{tab:noattack}}
%\begin{adjustwidth}{-\extralength}{0cm}
%\newcolumntype{C}{>{\centering\arraybackslash}X}
    \begin{tabular}{|c|c|c|c|c|}
    \hline
    \textbf{Model $\rightarrow$}    &   \textbf{MLP} & \textbf{DT} & \textbf{DT}    &   \textbf{XGBoost} \\ \hline \hline

\textbf{Accuracy $\rightarrow$} &   \cellcolor{green!94} 94     & \cellcolor{green!86}  86    &     \cellcolor{green!92}   92  &   \cellcolor{green!96} 96 \\
    \hline
    
    \end{tabular}
    %\end{adjustwidth}
\end{table}

\subsection{Baseline Models}
For comparison purpose, we first develop baseline four models using the above-mentioned algorithms without considering the attack. The accuracy of the four models on the test set are presented in table \ref{tab:noattack}. The results of these models are quite impressive as it is possible to make $95\%$ correct predictions with the default settings of the algorithms. We left the hyperparameter tuning of these algorithms to improve the performance further for our future work. 

\subsection{Attack and Defense Analysis}
In the analysis of the proposed attack and defense mechanism, in this paper, we seek answers to the following three interrelated questions which are critical in evaluating their effectiveness.
\begin{itemize}
    \item What impact does the proposed attack have on a HAR model?
    
    \item How good is the defense mechanism for a HAR model against the proposed attack?
    
    \item What is the influence of the amount of trusted data on the defense mechanism?
\end{itemize}
\subsubsection{{Impact of the Proposed Attack}}
To answer the first question, we study the relationship between the attack and different HAR models for varying amounts of poisoned data. The results on this relationship is showed in table \ref{tab:accu_attack}. With $1\%$ label flipped data, the attack barely has any significant affect on the four models. However, with little increase in the amount of poisoned data (from $1\%$ to $5\%$), the attack was able to influence all models' accuracy crucially. 
Among all the models, MLP's accuracy dropped drastically (from $94\%$ to $77\%$). Among the tree-based models, both random forest (RF) and XGboost show notable strength against the attack with $5\%$ poisoned data. However, their resistance against the attack is almost completely washed away under $15\%$ poisoned data. At this point, all the models become literally useless as the accuracy dropped to $50\%$ or less. 

\begin{table} 
\centering
\caption{Accuracy on the poisoned dataset (in \%)\label{tab:accu_attack}}
    \begin{tabular}{|c|c|c|c|c|c|c|c|}
    \hline
    
    \multirow{2}{*}{\textbf{Models}}  &  \multicolumn{7}{c|}{\textbf{Amount of Label Flipped Data (in \%)}} \\  \cline{2-8} 
    & 1 & 5 & 10 & 15 & 20 & 25 & 30 \\ \hline\hline
    
    MLP & \cellcolor{green!93}  93    & \cellcolor{green!77}  77    & \cellcolor{green!71} 71    &  \cellcolor{green!39} 39    & \cellcolor{green!27}  27    & \cellcolor{green!21}    21  & \cellcolor{green!19}  19 \\ \hline
    
    DT  &  \cellcolor{green!83} 83    &\cellcolor{green!65}  65    &  \cellcolor{green!63} 63    & \cellcolor{green!35} 35     & \cellcolor{green!27} 27    & \cellcolor{green!21} 21  & \cellcolor{green!18}  18  \\ \hline
    
    RF & \cellcolor{green!92}  92   & \cellcolor{green!84} 84   & \cellcolor{green!80} 80   & \cellcolor{green!56} 56     & \cellcolor{green!45}  45   & \cellcolor{green!35}35  &  \cellcolor{green!27} 27 \\ \hline
    
    XGBoost  &  \cellcolor{green!93} 93   & \cellcolor{green!83} 83   & \cellcolor{green!80}  80   &  \cellcolor{green!55} 55    & \cellcolor{green!45} 45    & \cellcolor{green!35}35  & \cellcolor{green!27} 27 \\ \hline
    
    \end{tabular}
    %\end{adjustwidth}
\end{table}

\begin{table} 
\centering
\caption{Accuracy on the recovered dataset after applying the defense mechanism with K = 9 (in \%)\label{tab:accu_defense}}
%\begin{adjustwidth}{-\extralength}{0cm}
%\newcolumntype{C}{>{\centering\arraybackslash}X}
    \begin{tabular}{|c|c|c|c|c|c|c|c|}
    \hline
    
     \multirow{2}{*}{\textbf{Models}}  &  \multicolumn{7}{c|}{\textbf{Amount of Label Flipped Data (in \%)}} \\  \cline{2-8} 
    & 1 & 5 & 10 & 15 & 20 & 25 & 30 \\ \hline\hline
    
    MLP & \cellcolor{green!94}  94   & \cellcolor{green!93}  93    & \cellcolor{green!93} 93    &  \cellcolor{green!91} 91    & \cellcolor{green!91}  91    & \cellcolor{green!89}    89  & \cellcolor{green!89}  89 \\ \hline

    DT  &  \cellcolor{green!85} 85    &\cellcolor{green!85}   85    &  \cellcolor{green!83} 83    & \cellcolor{green!84} 84     & \cellcolor{green!82} 82    & \cellcolor{green!82} 82  & \cellcolor{green!82}  82  \\ \hline
    
    RF & \cellcolor{green!92}  92   & \cellcolor{green!92} 92   & \cellcolor{green!91} 91   & \cellcolor{green!89} 89     & \cellcolor{green!90}  90   & \cellcolor{green!89}89  &  \cellcolor{green!88} 88 \\ \hline
    
    XGBoost  &  \cellcolor{green!94} 94   & \cellcolor{green!93} 93   & \cellcolor{green!92}  92   &  \cellcolor{green!90} 90    & \cellcolor{green!90} 90    & \cellcolor{green!90}90  & \cellcolor{green!88} 88 \\ \hline
    
    \end{tabular}
    %\end{adjustwidth}
\end{table}

\subsubsection{{Effectiveness of the Defense Mechanism against the Attack}}
\begin{figure*}
\centering
\begin{tabular}{c c c}
\hline
\cellcolor{blue!20}\textbf{Original Dataset}    &   \cellcolor{blue!20}\textbf{Poisoned Dataset}    & \cellcolor{blue!20}\textbf{Recovered Dataset} \\ \hline \hline\\
\includegraphics[width=0.27\linewidth]{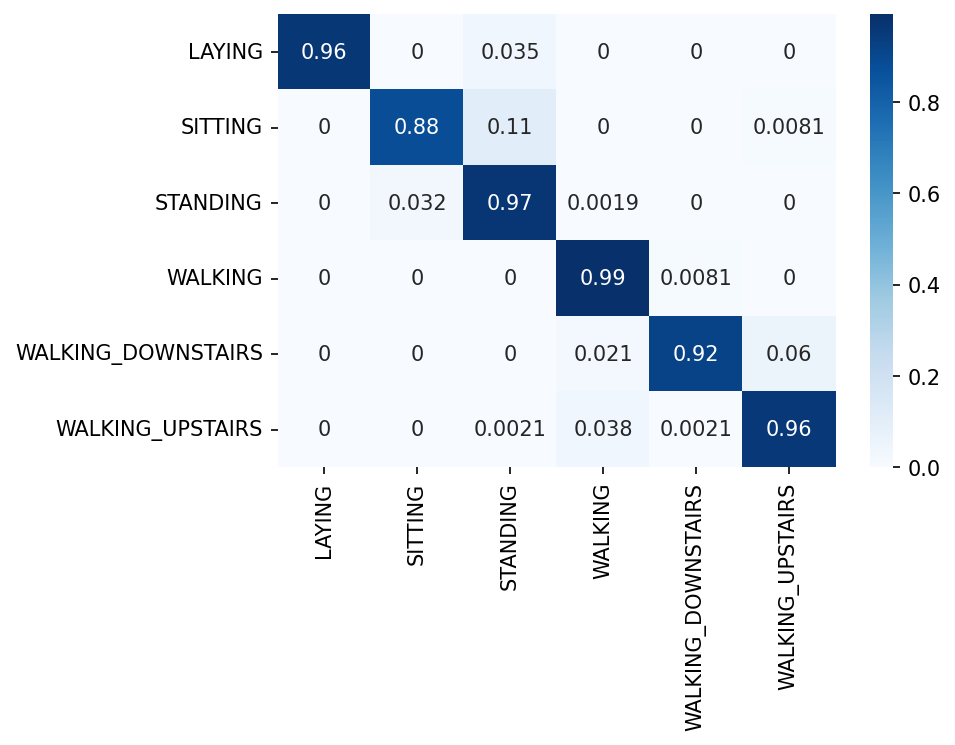}&
\includegraphics[width=0.27\linewidth]{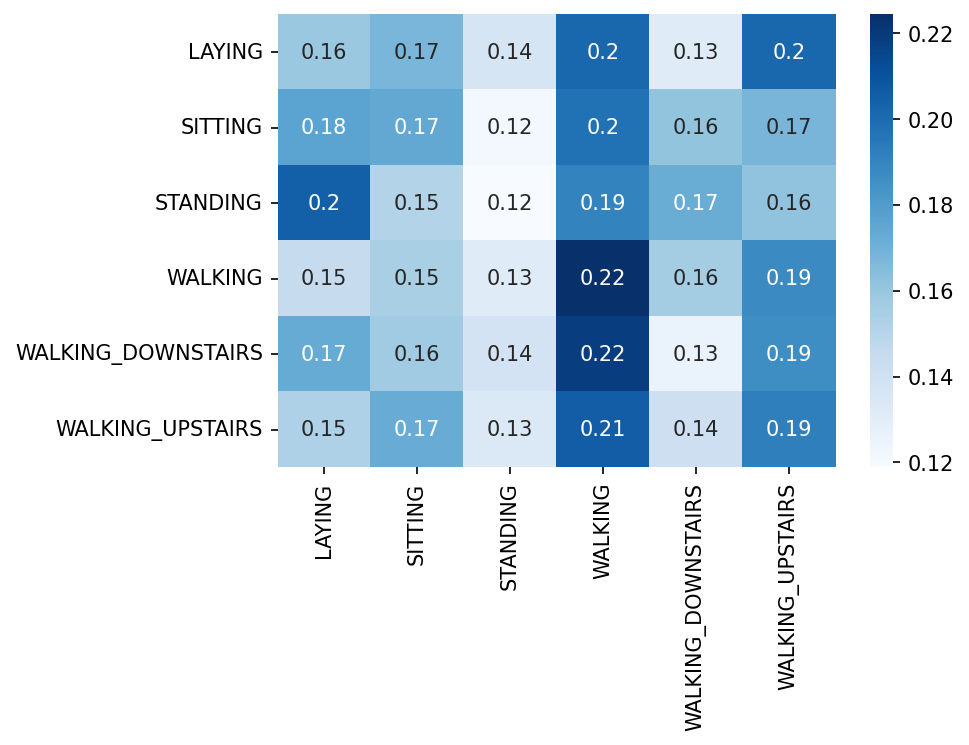}&
\includegraphics[width=0.27\linewidth]{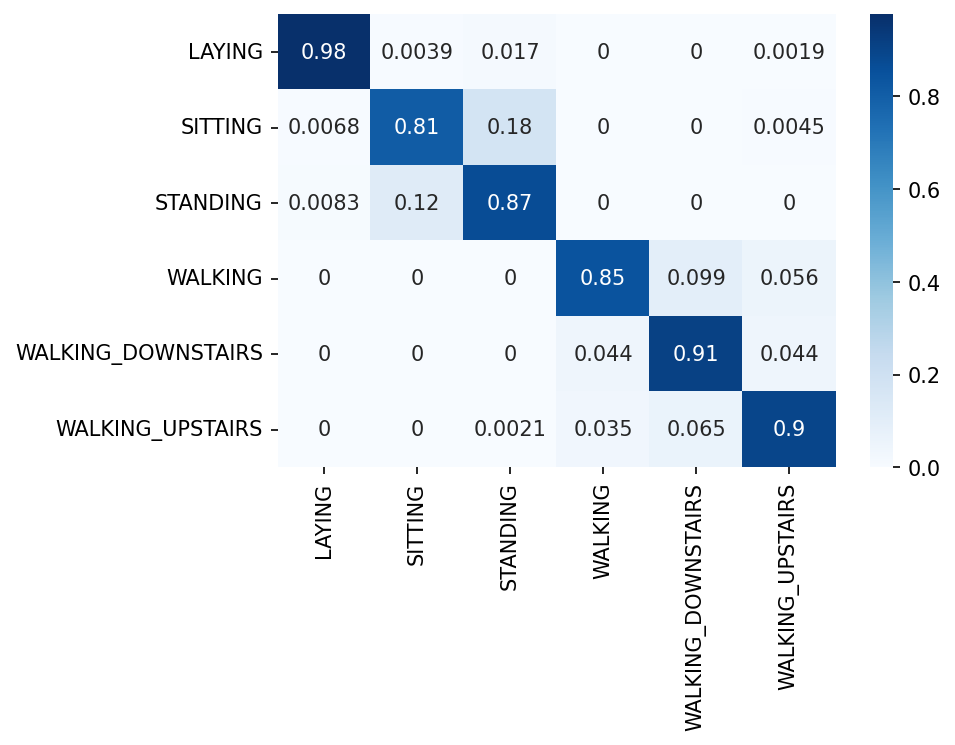}
\\ 
\multicolumn{3}{c}{Multi-layer Perceptron} \\ \hline\\

\includegraphics[width=0.27\linewidth]{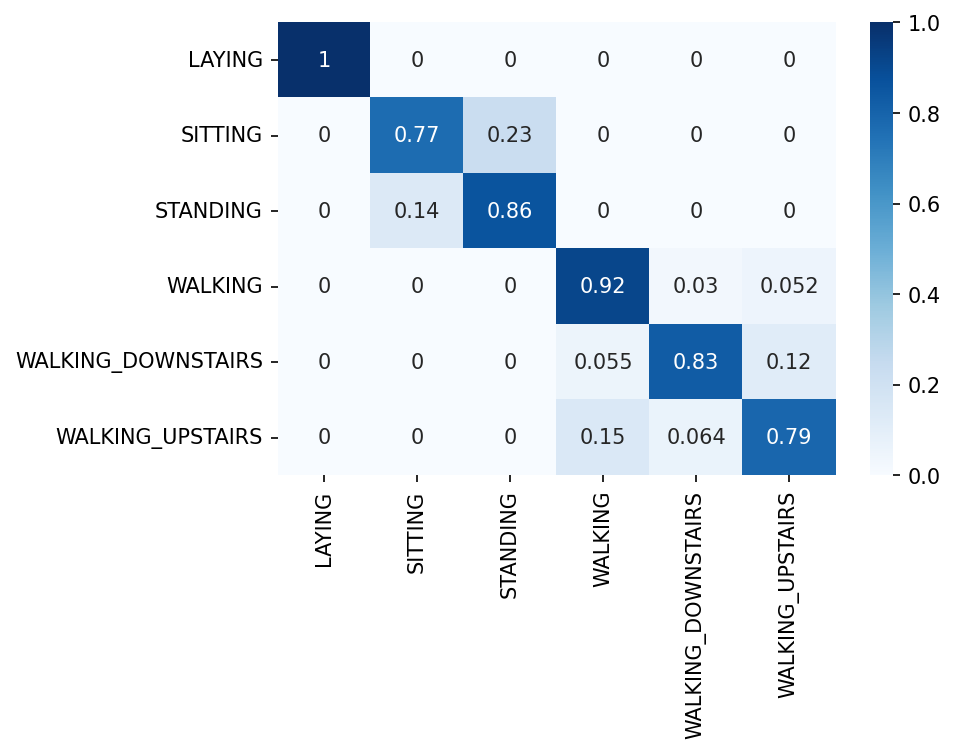}&
\includegraphics[width=0.27\linewidth]{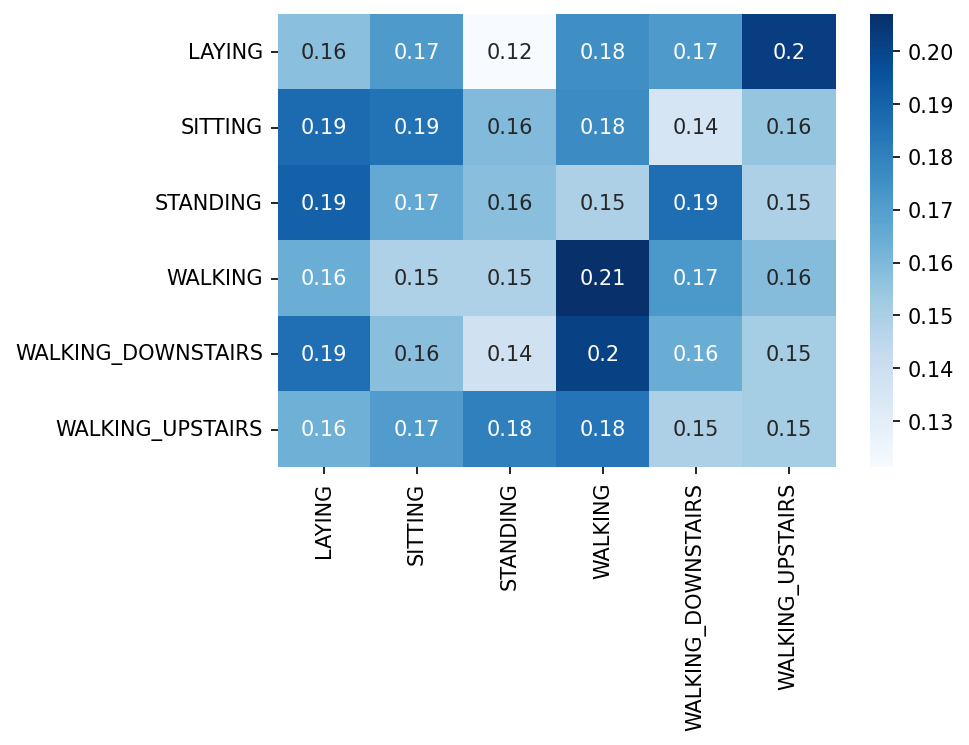}&
\includegraphics[width=0.27\linewidth]{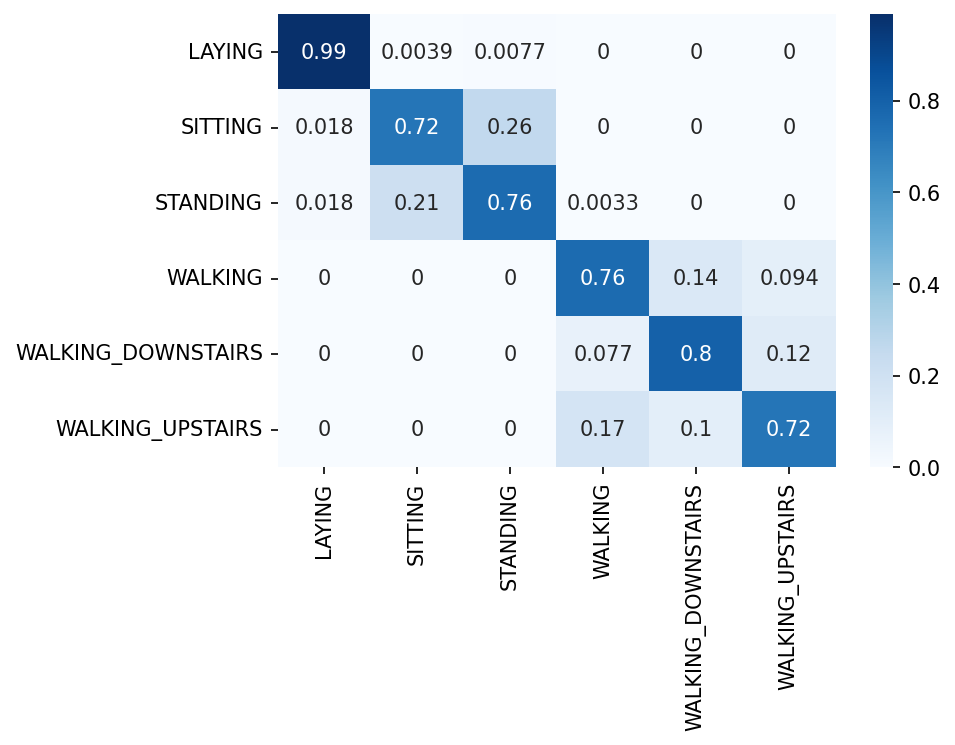} 
\\ 
\multicolumn{3}{c}{Decision Tree} \\ \hline\\
\includegraphics[width=0.27\linewidth]{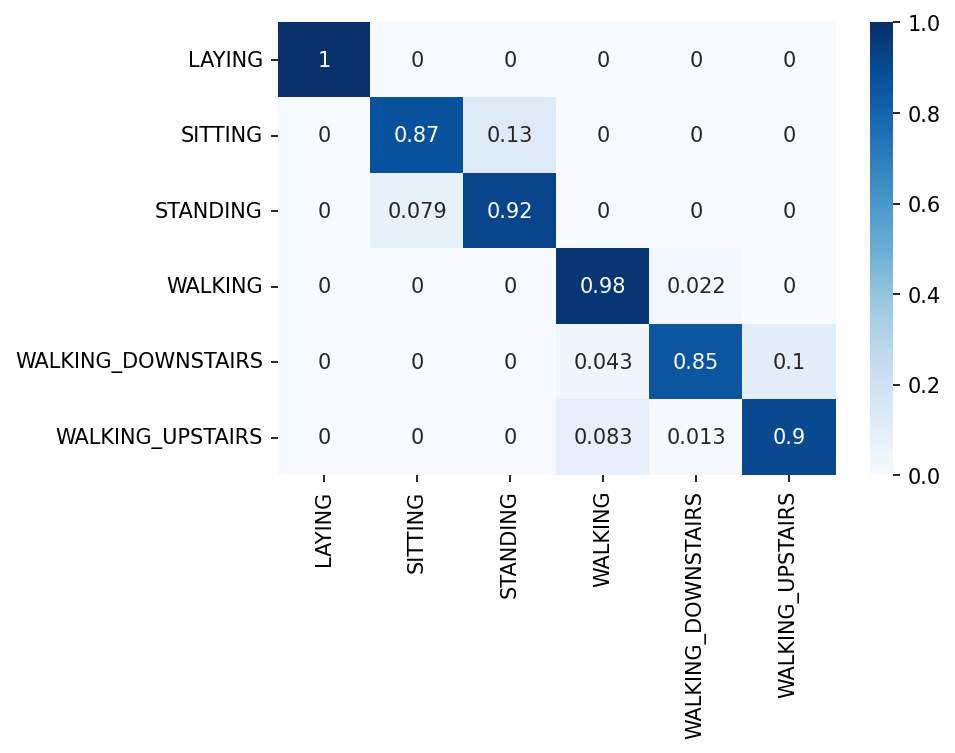}&
\includegraphics[width=0.27\linewidth]{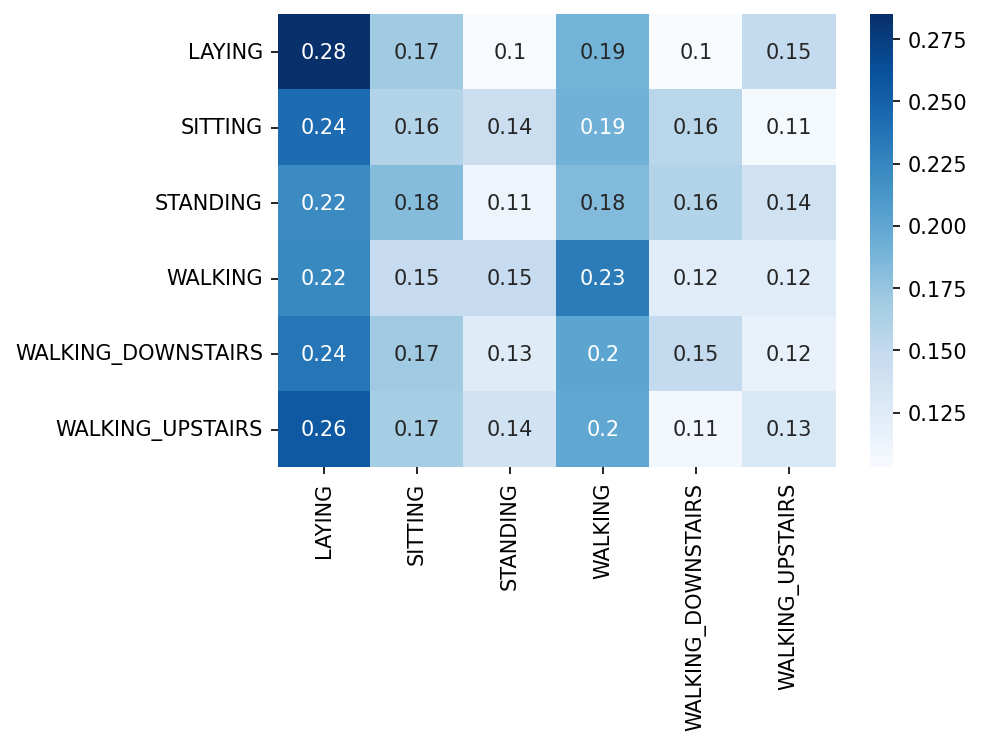}&
\includegraphics[width=0.27\linewidth]{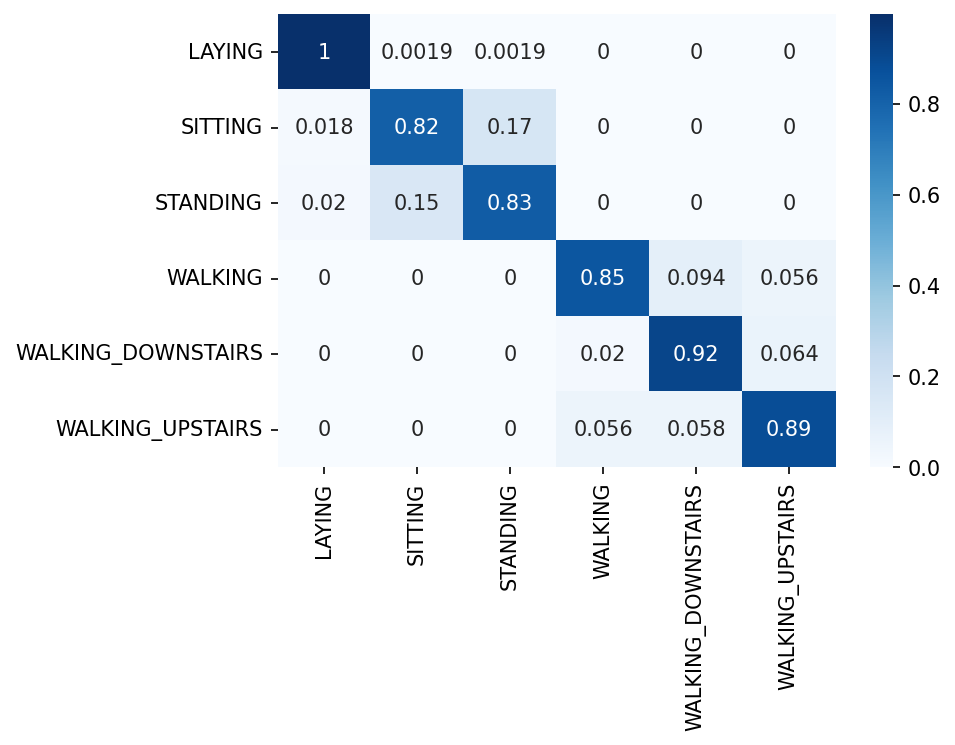}\\ 
\multicolumn{3}{c}{Random Forest} \\ \hline\\
\includegraphics[width=0.27\linewidth]{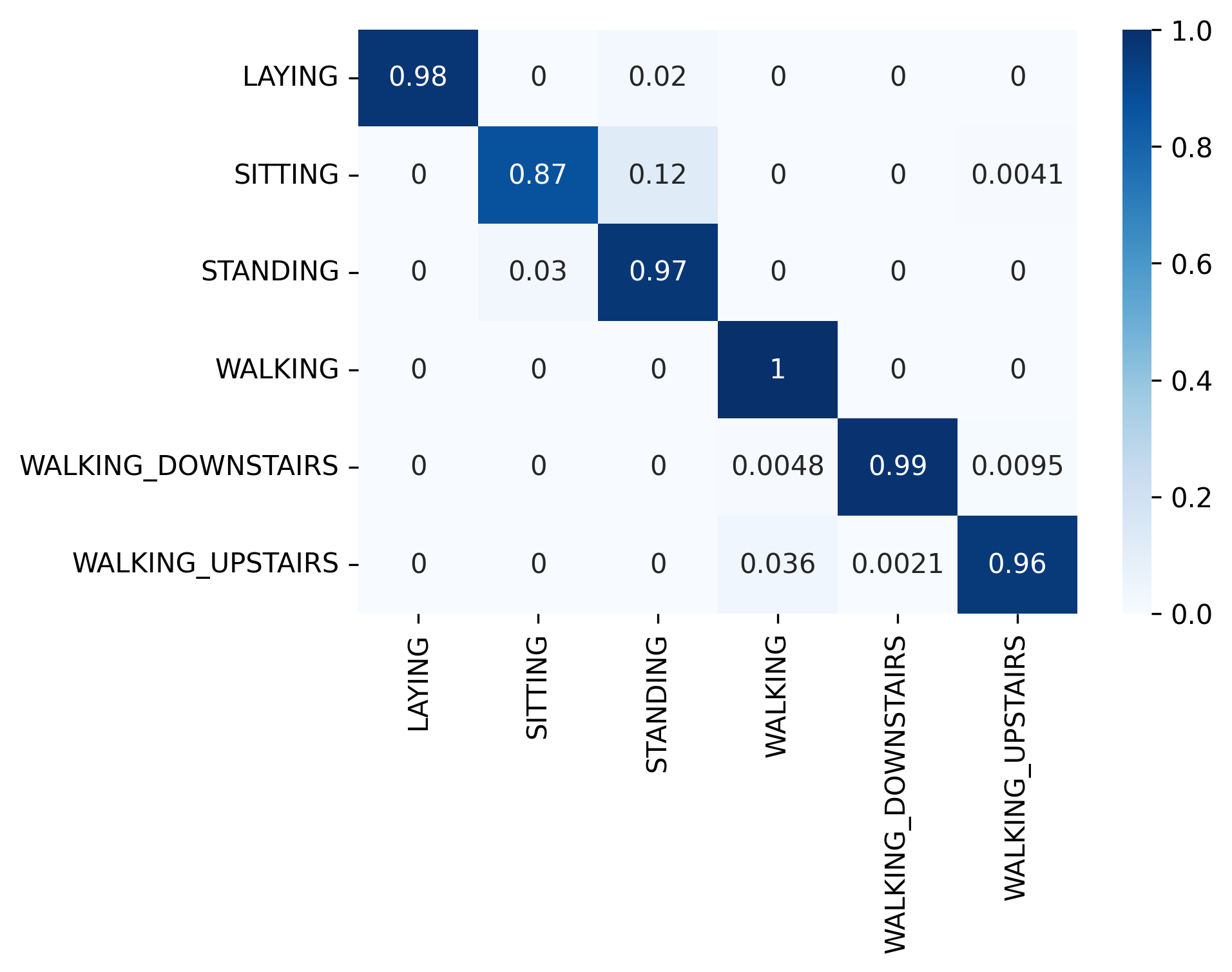}&
\includegraphics[width=0.27\linewidth]{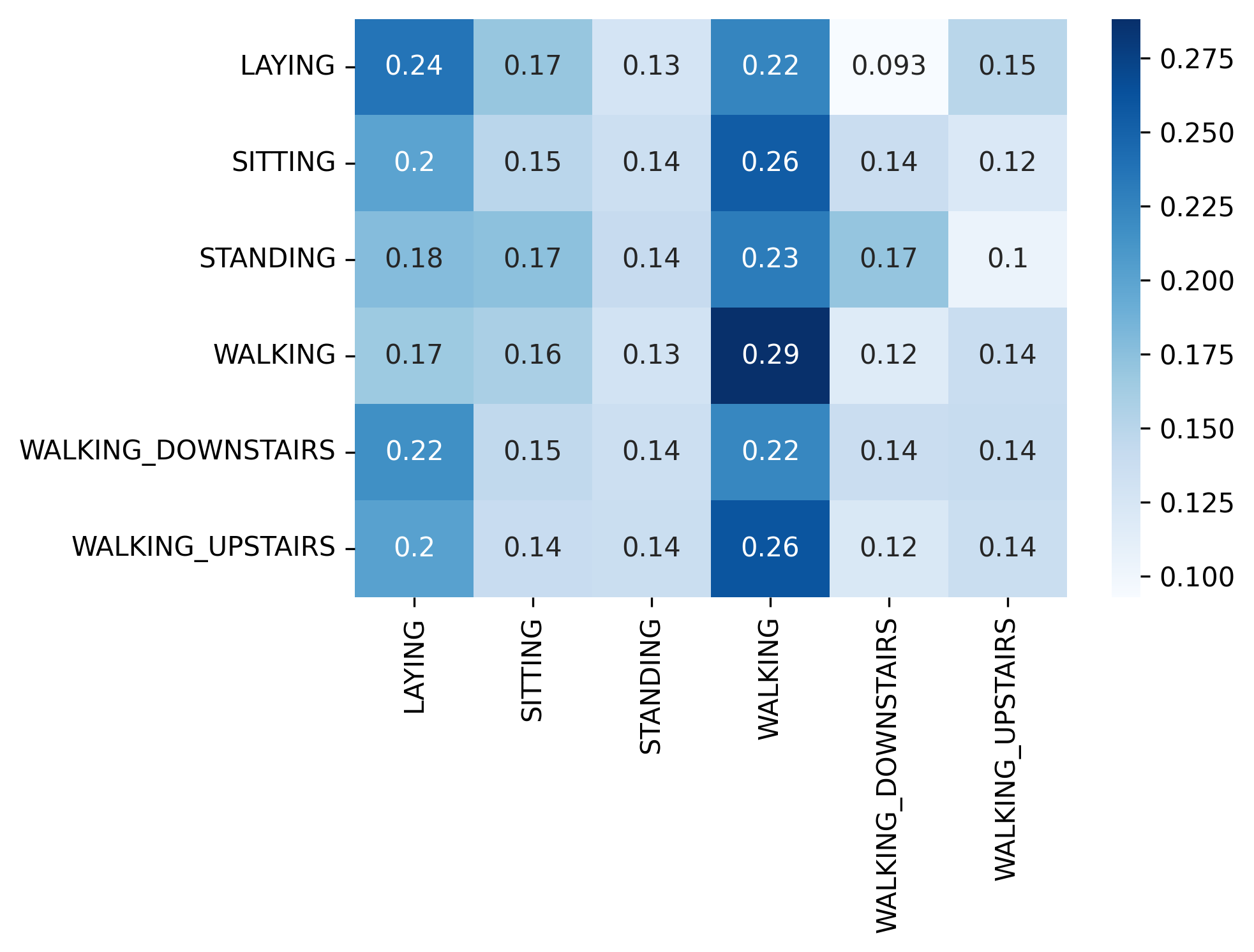}&
\includegraphics[width=0.27\linewidth]{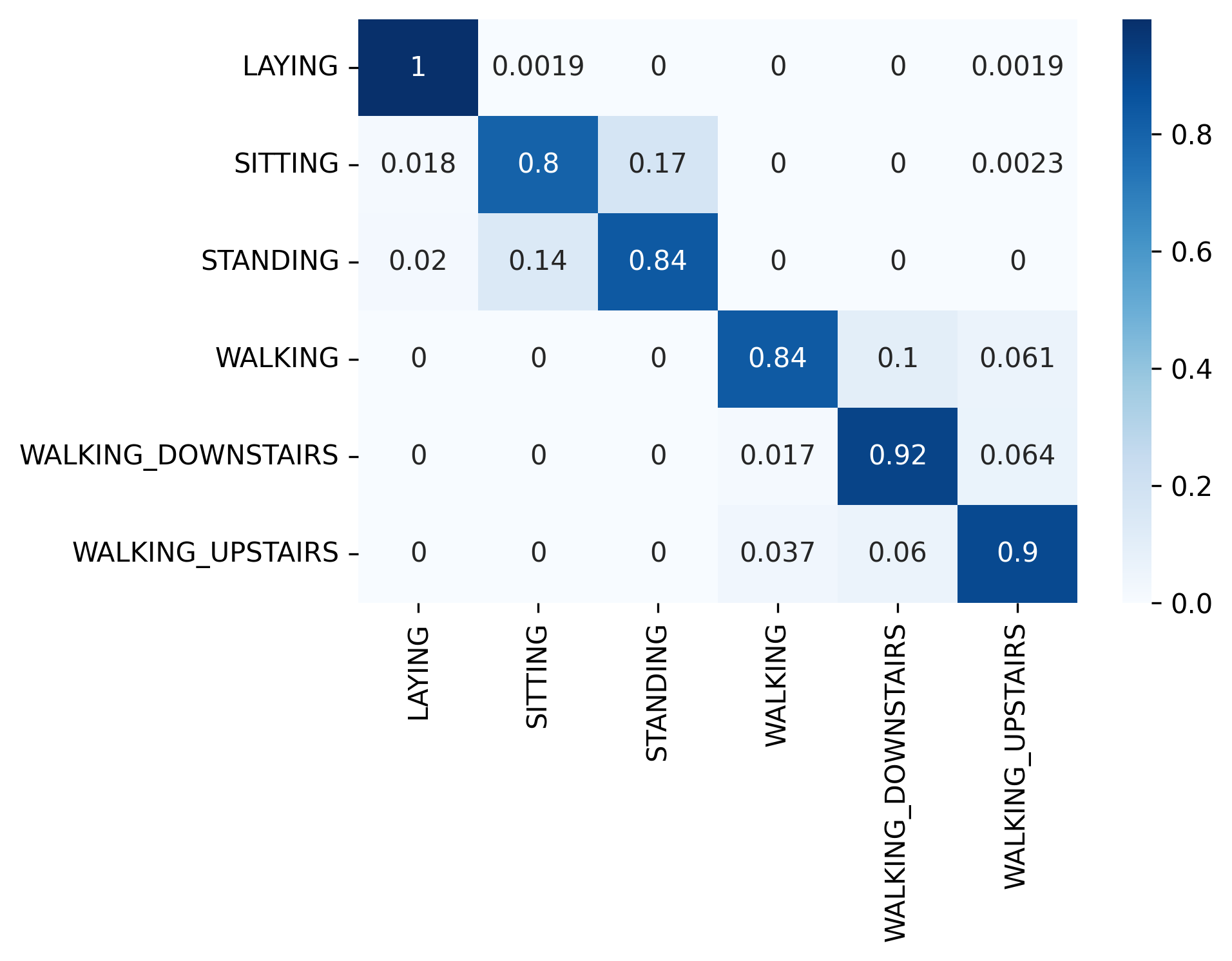}\\ 

\multicolumn{3}{c}{XGBoost} \\  \hline

\end{tabular}
%\end{adjustwidth}
\caption{Confusion matrix of the decision tree, random forest, and XGBoost models on the original, poisoned, and recovered datasets. Here, 10\% of the entire dataset was used as a trusted set. 25\% of the untrusted dataset was contaminated with the proposed label flipped attack. k = 9 was used to train the KNN-based defense model.}
\label{fig:confusion}
\end{figure*}

Next, we analyze the strength of the defense mechanism against the attack with varying amounts of poisoned data samples. We first examine its impact on the accuracy of the different models after sanitizing the poisoned data set using it (with $K=9$) in table \ref{tab:accu_defense}. Let us take a look at the scenario with $15\%$ poisoned data. The defense model was able to recover most of the poisoned data as the accuracy of the model increased sharply, and almost reached the baseline accuracy. For instance, while MLP's accuracy dropped to $39\%$ for $15\%$ poisoned data, after sanitizing the data with the defense model, it was able to achieve $91\%$ accuracy. Impressively, the models achieved pretty good accuracy even under a high volume of poisoned data (e.g. $30\%$). This signifies the effectiveness of the proposed KNN-based defense mechanism in sanitizing the label flipped data.  To get a better idea of what the models are getting right and what types of errors they are making on the original, poisoned, and recovered datasets, let us take a look at the confusion metrics of the different models presented in figure \ref{fig:confusion}. Note that these confusion matrices were generated with $K=9$, $x = 10\%$, $p = 25\%$. The first insight we can draw from the confusion metrics on the poisoned dataset is that the attack exhibits an untargeted attack as it scrambled all the numbers in each matrix. Another insight that can be drawn is the similarities between the confusion matrices on original and recovered datasets for all the models. Apparently, all the models had some difficulties in differentiating SITTING and STANDING; and it is not due to the attack, as a similar problem was perceived on the original dataset too. One potential solution to this problem is to collect more data on these two activities.

\subsubsection{Influence of Trusted Dataset Size on the Defense Mechanism}
\begin{figure}
%\begin{adjustwidth}{-\extralength}{0cm}
\centering
\includegraphics[width=\linewidth]{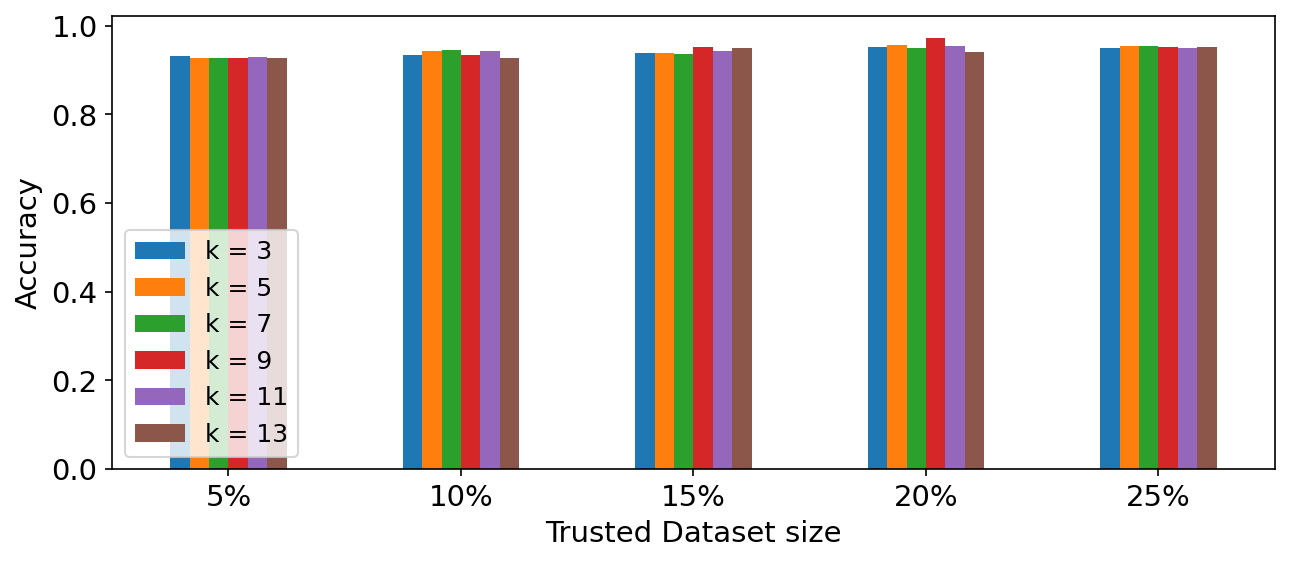}
%\end{adjustwidth}
\caption{Relationship between the defense model and trusted dataset size \label{fig:trustedset}}
\end{figure}

Figure \ref{fig:trustedset} presents experimental results of the relationship between the size of the trusted dataset and the KNN-based defense mechanism. Here, by accuracy, we mean the accuracy of the mechanism with a specific value of $K$ to correctly predict the label of a sample in a trusted dataset. For all the $K$s, regardless the size of the trusted dataset, the models achieved more than $90\%$ accuracy. In addition, the models show improved performance with the increase in the size of the trusted dataset.   

\section{Conclusion}
The open structure of the human activity recognition (HAR) system for data collection, such as mobile crowdsensing, allows adversaries to inject poisonous data to disrupt the recognition process. In this paper, we present our novel research on investigating the label flipping data poisoning attack on wearable-sensor-based HAR systems. While the majority of the attacks on label flipping focus on the binary classifier, our resea extend the existing works for multi-class classification problem. We empirically demonstrate the effectiveness of the proposed attack on different machine learning algorithms. Then, we present a defense mechanism based on the KNN algorithm which exhibits impressive performance on a real-world dataset. As an early work, the work has scope for improvement that lay the foundation for future research works, which includes, developing a better defense mechanism as the KNN-based approach has its limitations, covering different datasets in the experiment, and evaluating the strength of the attack against deep learning models.

\bibliographystyle{IEEEtran}
%\bibliography{bib_har.bib}
\bibliography{main.bbl}

\end{document}